\documentclass[twocolumn,showpacs,amsmath,amssymb,prc]{revtex4}
\usepackage{mathrsfs}

\usepackage{graphicx,color}
\usepackage{dcolumn}
\usepackage{bm}
\usepackage{CJK}

\begin{document}
\begin{CJK*}{GBK}{song}

\title{Improved radial basis function approach with the odd-even corrections}

\author{Z. M. Niu$^1$}\email{zmniu@ahu.edu.cn}
\author{B. H. Sun$^2$}
\author{H. Z. Liang$^{3,4}$}
\author{Y. F. Niu$^{5}$}
\author{J. Y. Guo$^1$}

\affiliation{$^1$School of Physics and Material Science, Anhui University,
             Hefei 230601, China}
\affiliation{$^2$School of Physics and Nuclear Energy Engineering, Beihang University,
             Beijing 100191, China}
\affiliation{$^3$RIKEN Nishina Center,
             Wako 351-0198, Japan}
\affiliation{$^4$Department of Physics, Graduate School of Science,
            The University of Tokyo, Tokyo 113-0033, Japan}
\affiliation{$^5$INFN, Sezione di Milano, I-20133 Milano, Italy}

\date{\today}

\begin{abstract}
The radial basis function (RBF) approach has been used to improve the mass predictions of nuclear models. However, systematic deviations exist between the improved masses and the experimental data for nuclei with different odd-even parities of ($Z$, $N$), i.e., the (even $Z$, even $N$), (even $Z$, odd $N$), (odd $Z$, even $N$), and (odd $Z$, odd $N$). By separately training the RBF for these four different groups, it is found that the systematic odd-even deviations can be cured in a large extend and the predictive power of nuclear mass models can thus be further improved. Moreover, this new approach can better reproduce the single-nucleon separation energies. Based on the latest version of Weizs\"{a}cker-Skyrme model WS4, the root-mean-square deviation of the improved masses with respect to known data falls to $135$~keV, approaching the chaos-related unpredictability limit ($\sim 100$ keV).
\end{abstract}

 \maketitle

\section{Introduction}

Nuclear mass is a basic quantity in nuclear physics, since it plays an important role not only in nuclear physics~\cite{Lunney2003RMP}, but also in other branches of science, such as astrophysics and cosmology~\cite{Burbidge1957RMP}. During the past decades,  great progress has been made in the mass measurements of atomic nuclei. More than $2000$ nuclear masses have been determined precisely~\cite{Wang2012CPC}. By analyzing the systematic trends in the mass surface and its derivatives, masses of nuclei near the region of known masses can be estimated with the Audi-Wasptra extrapolation method~\cite{Wang2012CPC, Audi2012CPC, Audi2003NPA, Audi1995NPA}. The local mass relations, such as the Garvey-Kelson (GK) relations~\cite{Garvey1966PRL, Garvey1969RMP, Barea2008PRC, Bao2013PRC, Cheng2014PRC}, the systematics of proton-neutron interactions~\cite{Zhang1989PLB, Fu2011PRC} and the Coulomb-energy displacement~\cite{Sun2011SCPMA, Kaneko2013PRL}, and systematics of $\alpha$-decay energies~\cite{Dong2011PRL, Li2013PRC} can also be applied to predict the masses of nuclei near the known region, although the intrinsic errors grow rapidly when these local mass relations are used for predictions in an iterative way.

To predict masses of nuclei far from known region, two types of global mass models are nowadays widely used: the macroscopic-microscopic and microscopic mass models. In the macroscopic-microscopic mass model, e.g., the finite-range droplet model (FRDM)~\cite{Moller1995ADNDT} and the Weizs\"{a}cker-Skyrme (WS) model~\cite{Wang2010PRCa}, the microscopic correction energies are extracted from the single-particle levels predicted with phenomenological potential, so numerical computations with this type of mass model is relatively fast. Moreover, various physical effects can be easily taken into account. For instance, by including the surface diffuseness correction for unstable nuclei, the accuracy of WS model has been significantly improved. The corresponding root-mean-square (rms) deviation with respect to data in the atomic mass evaluation of 2012 (AME2012) falls to $298$~keV, crossing the $0.3$~MeV accuracy threshold for the first time within the mean-field framework~\cite{Wang2014PLB}. Guiding by the semi-classical periodic orbit theory, a new macroscopic-microscopic mass model is proposed (for simplicity, this mass model is named as Bhagwat model hereafter), in which the fluctuating part of nuclear mass is taken as functions of the Fermi momentum of neutron and proton (i.e., $N^{1/3}$ and $Z^{1/3}$) and the proximity to shell closures~\cite{Bhagwat2014PRC}. The Bhagwat mass formula yields an rms deviation of $266$~keV, which is one of the smallest deviations reported in literatures to our knowledge.

In contrast to the macroscopic-microscopic model, the mass calculation with microscopic models is usually much complicated, but it is usually thought to have a better extrapolation ability. This kind of models has their own merits, e.g., it can be used to predict various nuclear properties within a unified framework, including not only the ground-state properties~\cite{Meng2006PPNP, Vretenar2005PRp, Meng2015JPG, Niu2013PLB, Niu2013PRCR} but also the excited-state properties~\cite{Meng2016Book, Niksic2011PPNP, Liang2013PRC}. Based on the Hartree-Fock-Bogoliubov (HFB) method with Skyrme or Gogny force, a series of microscopic mass models have been proposed~\cite{Goriely2009PRLa, Goriely2009PRLb, Goriely2013PRCR}. Apart from the non-relativistic ones, the relativistic mean-field (RMF) model has also been employed for the systematic calculations of nuclear masses recently~\cite{Geng2005PTP, Zhao2012PRC, Meng2013FP, Zhang2014FP}.

Although these global mass models have achieved great progress in recent years, the accuracies in their predictions are generally worse than those from the local mass formulas ($\sim 100$ keV). Meanwhile, the deviations among the mass predictions of different global models are still widely distributed over the nuclear chart. Recently, the CLEAN image reconstruction technique was applied to improve the accuracy of nuclear mass models, and indeed it significantly reduces the rms deviation to the known masses~\cite{Morales2010PRC}. An alternative simple and effective approach to improve the accuracy of mass models is the radial basis function (RBF). Recent works~\cite{Wang2011PRC, Niu2013PRCb,Zheng2014PRC} already demonstrated its ability to improve the description of nuclear masses and the two-neutron severation energies. However, it was also noticed that remarkable odd-even staggering exists between the masses improved by the RBF approach and the experimental data. The problem accordingly deteriorates the description of nuclear single-neutron separation energies~\cite{Zheng2014PRC}. In this work, we attempt to cure the odd-even deviations found in our previous RBF calculations, which will be presented in Sec.~III. For completeness, a brief introduction to the RBF approach will be given in Sec.~II, and a summary will be presented in Sec.~IV.

\section{Radial basis function approach and numerical details}

The RBF approach was first introduced to improve the mass predictions of nuclear models in Ref.~\cite{Wang2011PRC}. More details can be found in Ref.~\cite{Niu2013PRCb, Zheng2014PRC}. In this approach, the solution at point $x$ is represented as a sum of $m$ radial basis functions $\phi(\|x-x_i\|)$ weighted by an appropriate coefficient $\omega_i$, i.e.,
\begin{eqnarray}\label{Eq:Sx}
  S(x) = \sum_{i=1}^m \phi(\|x-x_i\|) \omega_i,
\end{eqnarray}
where $\|x-x_i\|$ is the Euclidean norm between point $x$ and center $x_i$. The weight $\omega_i$ is determined by training the RBF with $m$ samples $(x_i, d_i)$, which means the reconstructed function $S(x)$ at point $x_i$ is just the value $d_i$, i.e.,
\begin{eqnarray}\label{Eq:Dx}
\left(
  \begin{array}{c}
    d_1 \\
    d_2 \\
    ... \\
    d_m \\
  \end{array}
\right)
=
\left(
  \begin{array}{cccc}
    \phi_{11} & \phi_{12} & ... & \phi_{1m} \\
    \phi_{21} & \phi_{22} & ... & \phi_{2m} \\
    ...       & ...       & ... & ... \\
    \phi_{m1} & \phi_{m2} & ... & \phi_{mm} \\
  \end{array}
\right)
\left(
  \begin{array}{c}
    \omega_1 \\
    \omega_2 \\
    ... \\
    \omega_m \\
  \end{array}
\right),
\end{eqnarray}
where $\phi_{ij}=\phi(\|x_i-x_j\|)$ with $i, j=1, 2, ..., m$. Then the RBF weights are determined
to be
\begin{eqnarray}\label{Eq:Omega}
\left(
  \begin{array}{c}
    \omega_1 \\
    \omega_2 \\
    ... \\
    \omega_m \\
  \end{array}
\right)
=
\left(
  \begin{array}{cccc}
    \phi_{11} & \phi_{12} & ... & \phi_{1m} \\
    \phi_{21} & \phi_{22} & ... & \phi_{2m} \\
    ...       & ...       & ... & ... \\
    \phi_{m1} & \phi_{m2} & ... & \phi_{mm} \\
  \end{array}
\right)^{-1}
\left(
  \begin{array}{c}
    d_1 \\
    d_2 \\
    ... \\
    d_m \\
  \end{array}
\right).
\end{eqnarray}
Using the weights determined in Eq.~(\ref{Eq:Omega}), the reconstructed function $S(x)$ can be calculated with Eq.~(\ref{Eq:Sx}) for any point $x$.

For improving the mass predictions of nuclear models, the mass differences $D(Z,N)=M_{\textrm{exp}}(Z,N)-M_{\textrm{th}}(Z,N)$ between the experimental data $M_{\textrm{exp}}$ and those predicted with nuclear mass models $M_{\textrm{th}}$ are taken as the reconstructed samples. In addition, as in Refs.~\cite{Wang2011PRC, Niu2013PRCb, Zheng2014PRC}, the Euclidean norm is defined to be the distance between nuclei $(Z_i,N_i)$ and $(Z_j,N_j)$ on the nuclear chart:
\begin{eqnarray}
   r=\sqrt{(Z_i-Z_j)^2 + (N_i-N_j)^2},
\end{eqnarray}
and the basis function $\phi(r)=r$ is adopted. With Eq.~(\ref{Eq:Omega}), the weights are easily determined and consequently the reconstructed function $S(Z,N)$ for nucleus $(Z,N)$ can be calculated with Eq.~(\ref{Eq:Sx}). Then the revised mass for nucleus $(Z,N)$ is given by
\begin{eqnarray}
M_{\textrm{th}}^{\textrm{RBF}}(Z,N)=M_{\textrm{th}}(Z,N)+S(Z,N).
\end{eqnarray}

To evaluate the predictive power of nuclear mass models, the rms deviation, i.e.,
\begin{equation}
    \sigma_{\textrm{rms}}
   =\sqrt{\frac1n
    \sum_{i=1}^n(M^i_{\textrm{th}}-M^i_{\textrm{exp}})^2},
\end{equation}
is usually employed, where $M^i_{\textrm{th}}$ and $M^i_{\textrm{exp}}$ are the theoretical and
experimental nuclear masses, respectively, and $n$ is the number of nuclei contained in
a given set. In this work, only the nuclei with $N\geqslant 8$ and $Z\geqslant 8$ are involved and the experimental data are taken from AME2012~\cite{Audi2012CPC}, unless otherwise specified. For the theoretical mass models, we take the RMF~\cite{Geng2005PTP}, HFB-27~\cite{Goriely2013PRCR}, DZ10~\cite{Duflo1995PRC}, DZ31~\cite{Zuker2008RMF}, ETFSI-2~\cite{Goriely2000AIP}, FRDM~\cite{Moller1995ADNDT}, KTUY~\cite{Koura2005PTP}, WS4~\cite{Wang2014PLB}, and Bhagwat ~\cite{Bhagwat2014PRC} mass models as examples, spanning from the macroscopic-microscopic to microscopic models. For simplicity, nuclear mass models improved by the RBF approach are denoted with Model+RBF.

\section{Results and discussion}

\begin{figure}
\includegraphics[width=7.5cm]{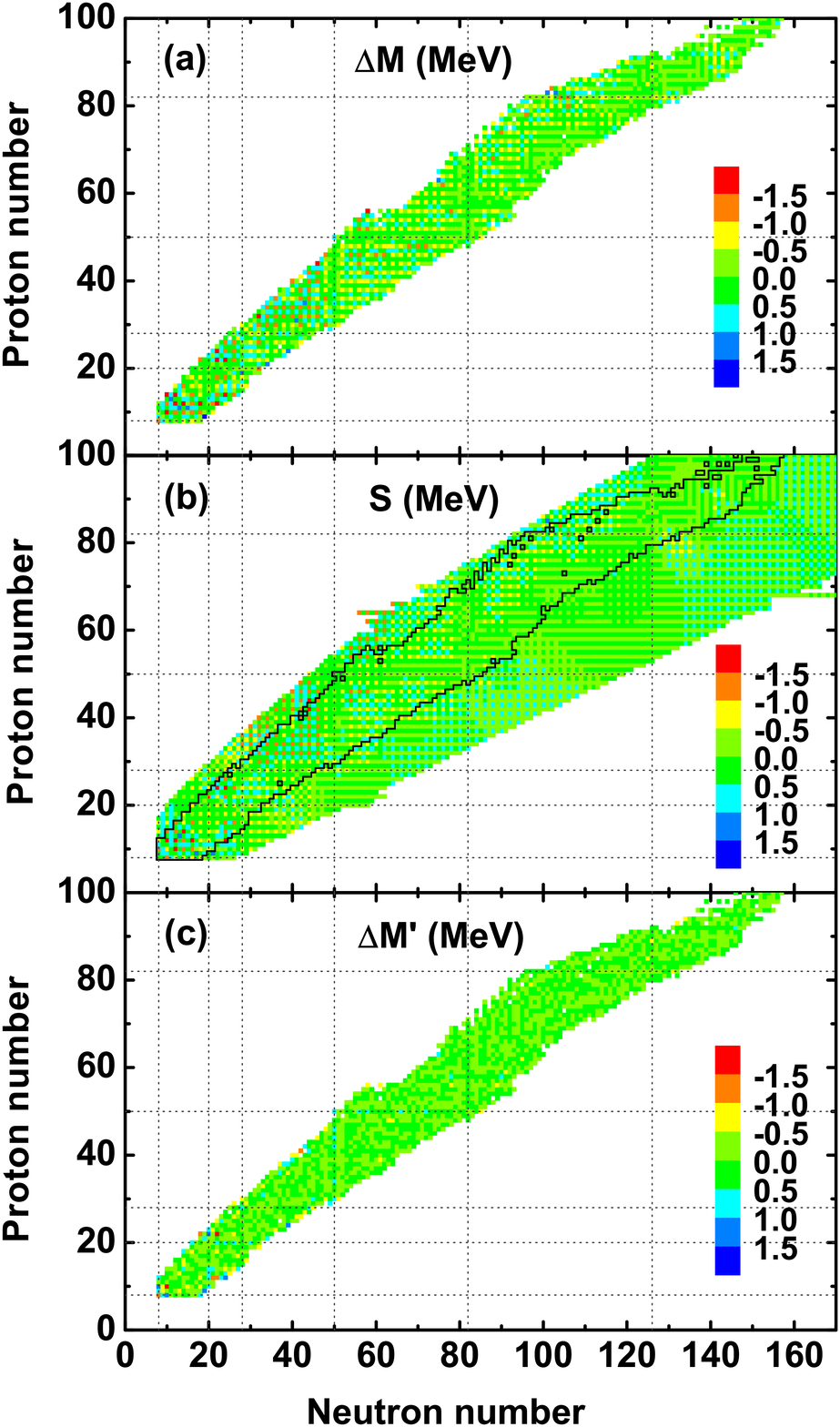}\\
\caption{(Color online) (a) Mass differences ($\Delta M$) between the experimental data and the predictions of RMF+RBF model. (b) The reconstructed functions $S(N,Z)$ predicted by the RBFoe approach. The boundary of nuclei with known masses in AME2012 is shown by the black contours. (c) Mass differences ($\Delta M'$) between experimental data and the predictions of RMF+RBFoe model. The dotted lines denote the conventional magic numbers.}\label{fig3}
\end{figure}

In previous studies, the reconstructed function $S(N,Z)$ for a known nucleus is calculated based on the remaining nuclei with known masses in AME2012~\cite{Niu2013PRCb, Zheng2014PRC}, i.e., the leave-one-out cross-validation method. As indicated in panel (c) of Fig.~1 in Ref.~\cite{Zheng2014PRC},  the differences between the masses improved by the RBF approach and the experimental data show remarkable odd-even staggering. For convenience, we redraw this plot in panel (a) of Fig.~\ref{fig3}. Although the strengths of such an odd-even staggering differ from region to region, it does exist in the whole nuclear chart.

\begin{figure}
\includegraphics[width=9cm]{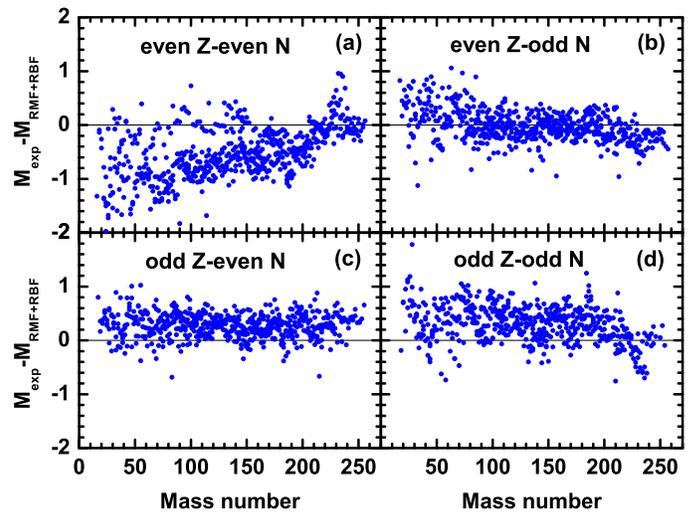}\\
\caption{(Color online) Mass differences between the experimental data and the predictions of RMF+RBF model for the (a) e-e, (b) e-o, (c) o-e, and (d) o-o nuclei.}\label{fig1}
\end{figure}

To better investigate this odd-even staggering, first of all, it would be instructive to organize the data into 4 groups, characterized by the odd-even parity of ($Z$, $N$), i.e., the (even $Z$, even $N$), (even $Z$, odd $N$), (odd $Z$, even $N$), and (odd $Z$, odd $N$). The mass differences between the experimental data and the predictions of the RMF+RBF model for the even-$Z$-even-$N$ (e-e) nuclei, even-$Z$-odd-$N$ (e-o) nuclei, odd-$Z$-even-$N$ (o-e) nuclei, and odd-$Z$-odd-$N$ (o-o) nuclei are shown in panels (a), (b), (c), and (d) of Fig.~\ref{fig1}, respectively. The deviations differ significantly group by group. Specifically, the masses are systematically overestimated for the e-e nuclei , while they are systematically underestimated for the o-e and o-o nuclei. For the e-o nuclei, the masses are underestimated in the light mass region with $A\lesssim 70$, while they are overestimated in the heavy mass region with $A\gtrsim 200$. This implies that the odd-even effect, originating from the quantum effect of the last few valence nucleons, is not able to be fully washed out when training the whole nuclear database simultaneously in a single RBF approach.

\begin{figure}
\includegraphics[width=9cm]{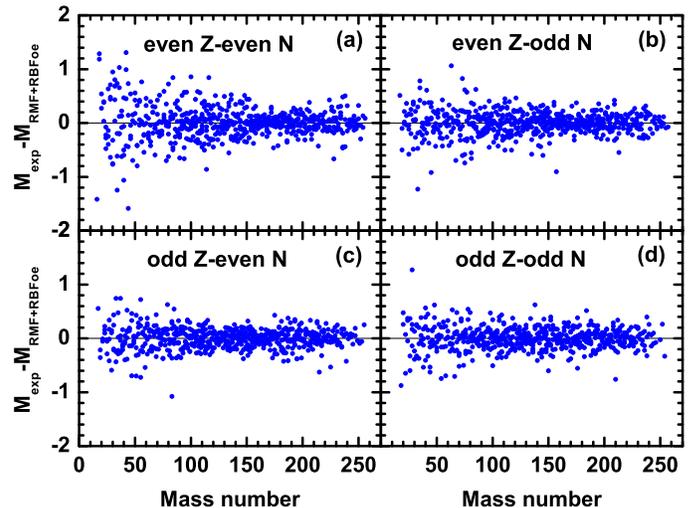}\\
\caption{(Color online) Same as Fig.~\ref{fig1}, but for the RMF+RBFoe model.}\label{fig2}
\end{figure}

Since the RBF approach can well eliminate the local systematic deviations, these systematic deviations for the e-e, e-o, o-e, and o-o nuclei should be removed with the RBF approach if we train the RBF separately for these four groups. For simplicity, such an RBF approach considering the odd-even effects is called to be the RBFoe approach, and the mass models improved by the RBFoe approach is denoted by Model+RBFoe hereafter. In panels (a), (b), (c), and (d) of Fig.~\ref{fig2}, the mass differences between the experimental data and the predictions of the RMF+RBFoe model are shown for the e-e, e-o, o-e, and o-o nuclei, respectively. As we expect, the systematic deviations for these four groups are well removed by the RBFoe approach in the whole mass region, although there are relatively larger scattering in the light mass region.

To understand the effects of the RBFoe approach, the reconstructed functions $S(N,Z)$ predicted by the RBFoe approach and the mass differences between the experimental data and the predictions of RMF+RBFoe approach are shown in panels (b) and (c) of Fig.~\ref{fig3}, respectively. The mass differences between the experimental data and the predictions of RMF+RBF approach are shown in panel (a) for comparison. It is found that the odd-even staggering shown in panel (a) can be well reproduced with the reconstructed functions inside the contour lines of panel (b). When the reconstructed functions are added to the mass predictions of RMF+RBF model, the odd-even staggering of mass deviations is well eliminated and the deviations are almost within $0.5$~MeV. From the structure outside the contour lines shown in Fig.~\ref{fig3}(b), it is found that the RBFoe approach predicts the odd-even corrections for the RMF+RBF model are within the $1$~MeV for the neutron-rich nuclei, while it is relatively larger for the neutron-deficient nuclei but still within $1.5$~MeV.

\begin{table*}
\begin{center}
\caption{The rms deviations (in MeV) of nuclear masses predicted by the Model+RBF and Model+RBFoe models with respect to the known masses in AME2012. The second (seventh) to sixth (eleventh) columns correspond respectively to the rms deviations of Model+RBF (Model+RBFoe) model for the e-e, e-o, o-e, and o-o nuclei, as well as all nuclei in AME2012 with $Z, N\geqslant8$.} \label{tb1}
\begin{tabular}{lcccccccccccccc}
\hline \hline
         &~  &\multicolumn{6}{c}{Model+RBF}                            &~~~   &\multicolumn{5}{c}{Model+RBFoe}                         \\
\hline
         &~  &e-e     &e-o     &o-e     &o-o     &~   &\textbf{total}  &~~~   &e-e     &e-o     &o-e     &o-o     &~   &\textbf{total} \\
\hline
RMF      &~  &0.718   &0.293   &0.358   &0.460   &~   &\textbf{0.488}  &~~~   &0.315   &0.223   &0.199   &0.219   &~   &\textbf{0.244} \\
HFB-27   &~  &0.328   &0.321   &0.322   &0.339   &~   &\textbf{0.328}  &~~~   &0.274   &0.272   &0.253   &0.216   &~   &\textbf{0.256} \\
DZ10     &~  &0.236   &0.220   &0.195   &0.247   &~   &\textbf{0.225}  &~~~   &0.153   &0.142   &0.137   &0.165   &~   &\textbf{0.150} \\
DZ31     &~  &0.223   &0.192   &0.166   &0.230   &~   &\textbf{0.204}  &~~~   &0.131   &0.135   &0.128   &0.173   &~   &\textbf{0.142} \\
ETFSI-2  &~  &0.431   &0.272   &0.405   &0.303   &~   &\textbf{0.360}  &~~~   &0.204   &0.159   &0.162   &0.188   &~   &\textbf{0.179} \\
FRDM     &~  &0.291   &0.255   &0.243   &0.282   &~   &\textbf{0.268}  &~~~   &0.221   &0.237   &0.221   &0.200   &~   &\textbf{0.221} \\
KTUY     &~  &0.238   &0.203   &0.181   &0.214   &~   &\textbf{0.210}  &~~~   &0.152   &0.144   &0.131   &0.149   &~   &\textbf{0.144} \\
WS4      &~  &0.198   &0.191   &0.180   &0.215   &~   &\textbf{0.196}  &~~~   &0.127   &0.129   &0.126   &0.158   &~   &\textbf{0.135} \\
Bhagwat  &~  &0.261   &0.196   &0.180   &0.238   &~   &\textbf{0.221}  &~~~   &0.158   &0.160   &0.149   &0.144   &~   &\textbf{0.153} \\
\hline \hline
\end{tabular}
\end{center}
\end{table*}

The RBFoe approach is then applied to various mass models, including the RMF, HFB-27, DZ10, DZ31, ETFSI-2, FRDM, KTUY, WS4, and Bhagwat models. The rms deviations of nuclear masses improved by the RBF and RBFoe approaches with respect to the known masses in AME2012 are given in Tab.~\ref{tb1} for these mass models. Clearly, the RBFoe approach significantly improves the mass descriptions of Model+RBF approach even separately for the e-e, e-o, o-e, and o-o nuclei. The rms deviations with respect to all the known masses in AME2012 are reduced by up to $50\%$ for the RMF and ETFSI-2 mass models. With the improvement of the RBFoe approach, all models considered here cross the $300$~keV accuracy threshold. Based on the latest version of WS model, WS4, the rms deviation falls to $135$~keV, even approaching the chaos-related unpredictability limit ($\sim 100$ keV) for the calculation of nuclear masses~\cite{Barea2005PRL}.

\begin{table*}
\begin{center}
\caption{The rms deviations (in MeV) of single-neutron separation energies ($S_n$), two-neutron separation energies ($S_{2n}$), single-proton separation energies ($S_p$), and two-proton separation energies ($S_{2p}$) with respect to the data in AME2012 for various mass models and their counterparts improved by the RBF and RBFoe approaches.} \label{tb2}
\begin{tabular}{lccccccccccccccc}
\hline \hline
          &~   &\multicolumn{4}{c}{Model}           &~~  &\multicolumn{4}{c}{Model+RBF}       &~~  &\multicolumn{4}{c}{Model+RBFoe}    \\
\hline
          &~   &$S_n$  &$S_{2n}$ &$S_p$  &$S_{2p}$  &~~  &$S_n$  &$S_{2n}$ &$S_p$  &$S_{2p}$  &~~  &$S_n$  &$S_{2n}$ &$S_p$  &$S_{2p}$ \\
\hline
RMF       &~   &0.648  &0.851    &0.809  &1.072     &~~  &0.661  &0.310    &0.799  &0.326     &~~  &0.333  &0.362    &0.333  &0.376    \\
HFB-27    &~   &0.425  &0.425    &0.434  &0.449     &~~  &0.523  &0.309    &0.494  &0.323     &~~  &0.385  &0.385    &0.372  &0.389    \\
DZ10      &~   &0.323  &0.420    &0.372  &0.493     &~~  &0.335  &0.194    &0.354  &0.235     &~~  &0.199  &0.231    &0.215  &0.267    \\
DZ31      &~   &0.287  &0.342    &0.304  &0.386     &~~  &0.325  &0.182    &0.322  &0.241     &~~  &0.203  &0.232    &0.222  &0.281    \\
ETFSI-2   &~   &0.438  &0.458    &0.486  &0.505     &~~  &0.532  &0.223    &0.569  &0.247     &~~  &0.236  &0.257    &0.248  &0.275    \\
FRDM      &~   &0.376  &0.493    &0.395  &0.502     &~~  &0.396  &0.271    &0.406  &0.274     &~~  &0.315  &0.333    &0.321  &0.335    \\
KTUY      &~   &0.307  &0.384    &0.358  &0.513     &~~  &0.320  &0.183    &0.302  &0.199     &~~  &0.197  &0.217    &0.206  &0.233    \\
WS4       &~   &0.258  &0.276    &0.274  &0.322     &~~  &0.300  &0.173    &0.290  &0.182     &~~  &0.189  &0.209    &0.192  &0.216    \\
Bhagwat   &~   &0.277  &0.270    &0.283  &0.294     &~~  &0.338  &0.187    &0.321  &0.197     &~~  &0.217  &0.233    &0.221  &0.230    \\
\hline \hline
\end{tabular}
\end{center}
\end{table*}

The odd-even staggering is directly related to nuclear single-nucleon separation energy, which is very important for nuclear reactions in astrophysics. Therefore, it is interesting to further investigate the description of nuclear single-nucleon separation energy for various mass models and their counterparts improved by the RBF and RBFoe approaches. The corresponding rms deviations of single-neutron separation energies $S_n$ and single-proton separation energies $S_p$ with respect to the data in AME2012 are given in Tab.~\ref{tb2}. For completeness, the description of two-neutron separation energies $S_{2n}$ and two-proton separation energies $S_{2p}$ are also shown. Comparing with original mass models, the RBF approach significantly improves the description of $S_{2n}$ and $S_{2p}$, whereas the description of $S_{n}$ and $S_{p}$ cannot be improved but even be deteriorated. With the RBFoe approach, the description of $S_{2n}$, $S_{2p}$, $S_{n}$, and $S_{p}$ are all significantly improved, although the description of $S_{2n}$ and $S_{2p}$ is slightly worse than the Model+RBF approach.

\begin{figure}
\includegraphics[width=7.5cm]{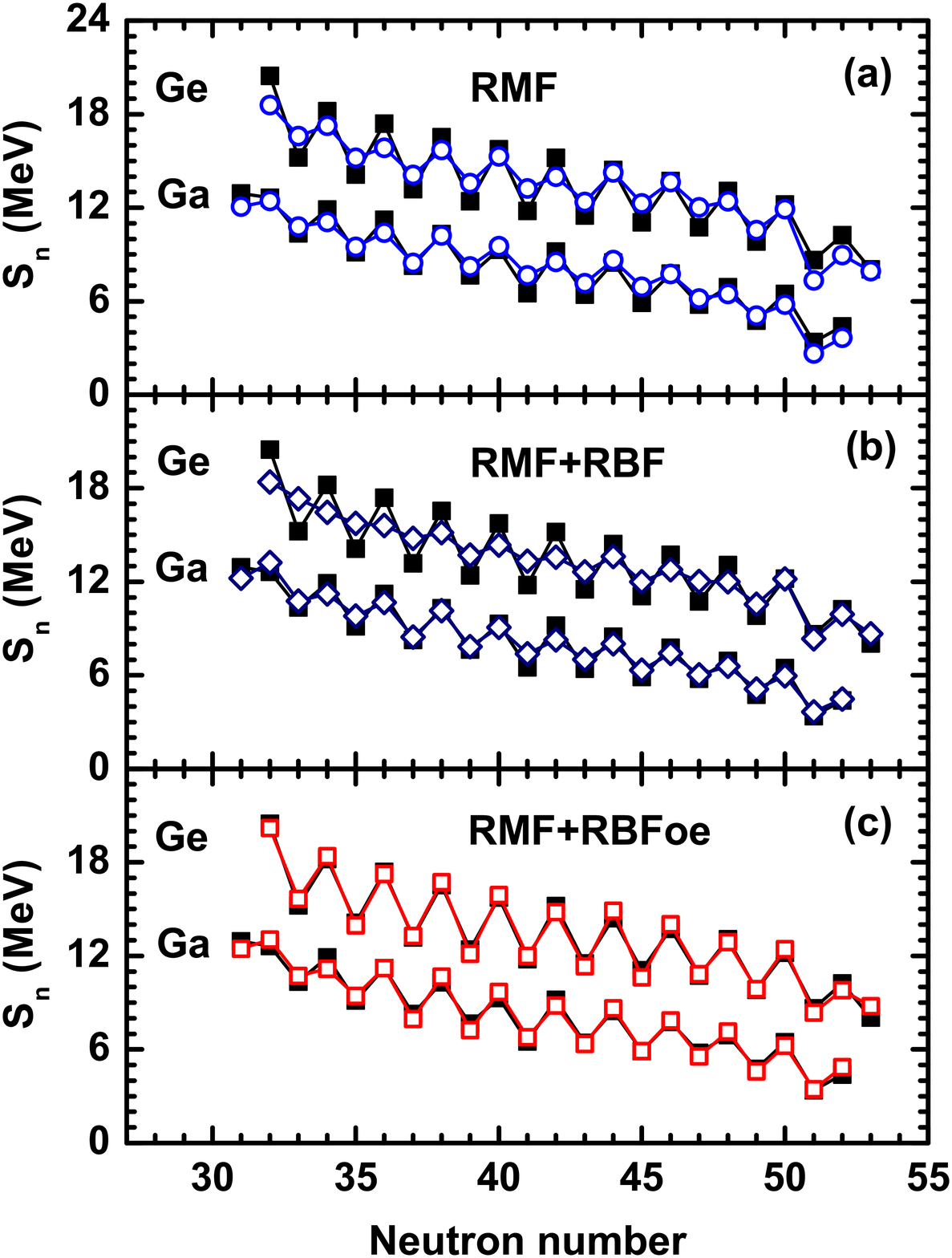}\\
\caption{(Color online) Single-neutron separation energies of the Ga and Ge isotopes predicted by (a) RMF, (b) RMF+RBF , and (c) RMF+RBFoe. The experimental data are shown with the filled squares. Note that $S_n$ of the Ge isotopes have been shifted up by $5$~MeV.}\label{fig4}
\end{figure}

For understanding how the RBFoe approach improves the description of nuclear single-nucleon separation energy, $S_{n}$ of the Ga (odd $Z$) and Ge (even $Z$) isotopes predicted by RMF are taken as examples. The corresponding results are shown in Fig.~\ref{fig4}. Note that $S_n$ of the Ge isotopes have been shifted up by $5$~MeV to avoid the mix with symbols for the Ga isotopes. Comparing with the RMF model, the RBF approach slightly improves the description of $S_n$ for the Ga isotopes, but the odd-even staggering predicted for the Ge isotopes is too weak to reproduce the experimental data. After considering the odd-even staggering, the RBFoe approach, on the other hand, gives an excellent description of $S_n$ for both Ga and Ge isotopes. This demonstrates the robustness of this new approach.

\begin{figure}[h]
\includegraphics[width=7.5cm]{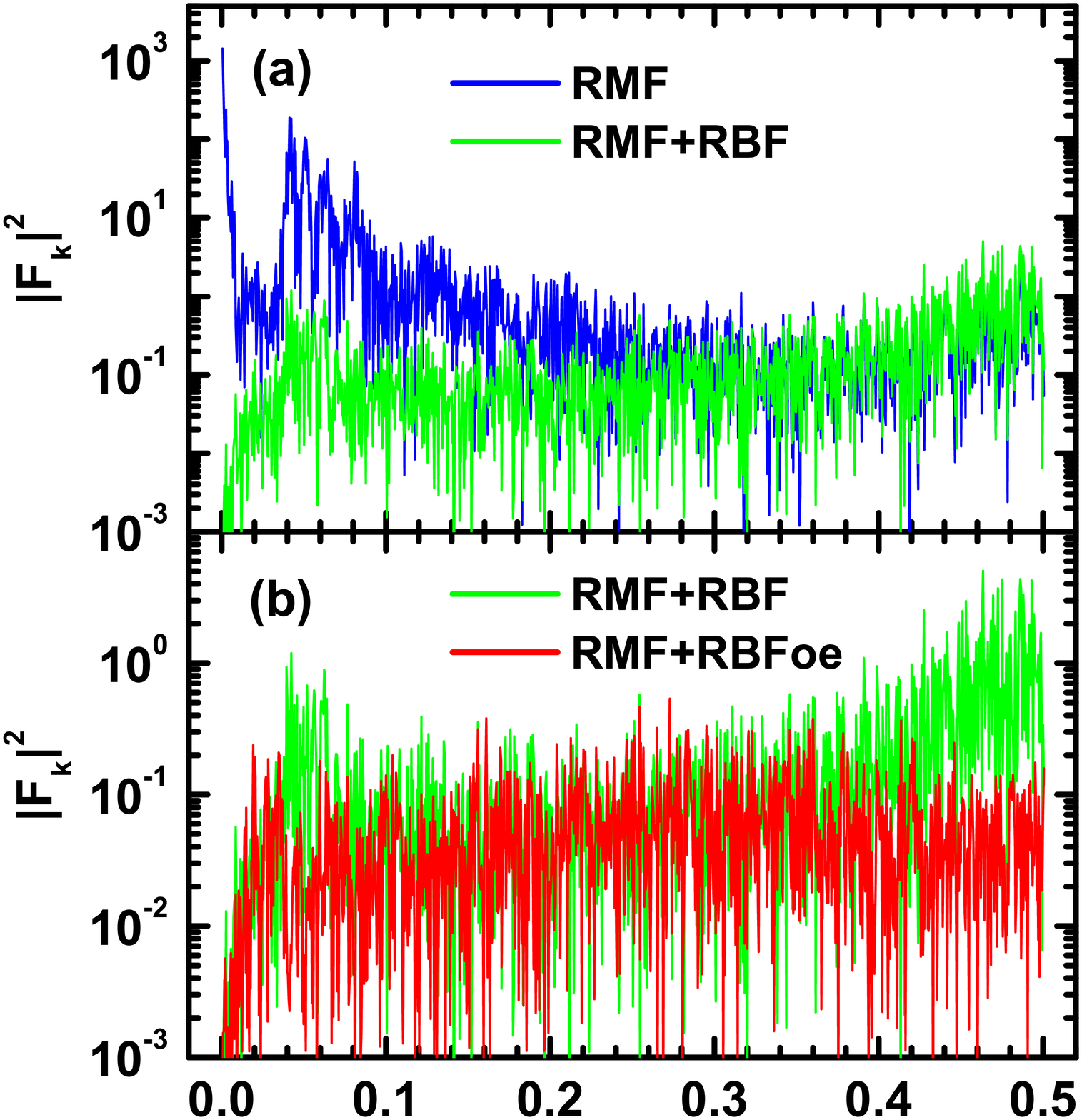}\\
\caption{(Color online) Squared amplitudes $|F_k|^2$ of the Fourier transforms of the mass differences as a function of the frequency $\omega=k/N$. For comparison, the results obtained for the RMF and RMF+RBF approaches are shown together in panel (a), and those obtained for the RMF+RBF and RMF+RBFoe approaches are shown together in panel (b).}\label{fig6}
\end{figure}

The Fourier transform of mass differences can give an insight for the correlations of mass differences. It is used to map similar trend of mass differences to similar frequency, i.e., the regularity in mass differences. Because the irregular form of the nuclear-data chart and a small number of nuclei along fixed $N$, $Z$, or $A$ lines, all nuclei with known masses are reordered as in Ref.~\cite{Barea2005PRL}: The nuclei are first ordered with increasing $A$. For the nuclei with the same $A$, they are further ordered by increasing $N-Z$ for the even-$A$ nuclei, while by decreasing $N-Z$ for the odd-$A$ nuclei. Then, the discrete Fourier transform of mass differences is evaluated with
\begin{equation}\label{Eq:Fk}
    F_k
   =\frac{1}{\sqrt{\mathcal{N}}} \sum_{j=1}^{\mathcal{N}} \frac{M_{\textrm{exp}}^j-M_{\textrm{th}}^j}{\gamma}
                              \exp\left[\frac{-2\pi i(j-1)(k-1)}{\mathcal{N}}\right]
\end{equation}
where $i$ is the imaginary unit, $j$ and $k$ are the order numbers, $\mathcal{N}$ is the total number of the mass data. The parameter $\gamma$ is included to make $F_k$ dimensionless. Since it only affects the global scale of the Fourier amplitudes, $\gamma=1$~MeV is adopted as in Ref.~\cite{Hirsch2004PRC}.

The squared amplitudes $|F_k|^2$ for the RMF, RMF+RBF, and RMF+RBFoe approaches are displayed in Fig.~\ref{fig6}. It should be noted that $|F_k|^2$ in Fig.~\ref{fig6} are shown in the logarithm scale. Very large components are found at low frequencies for the RMF model, which corresponds to the local and slow change of the mass differences of RMF model~\cite{Geng2005PTP}. With the RBF approach, this local mass differences are well eliminated, as reflected by the fact that the large $|F_k|^2$ values at low frequencies are significantly reduced. However, a slightly enhanced components are observed at high frequencies. From the preceding discussions, we already know that these high-frequency components mainly originate from the odd-even staggering of mass differences. These high-frequency components can be further reduced to a large extend with the RBFoe approach. No dominant large-frequency components are observed for the RBFoe approach, or in other words, it is difficult to find particular correlations of the mass differences over the nuclear chart. This again verifies the success of the RBFoe in reproducing the whole nuclear mass surface. Similar distributions of the Fourier amplitudes are also observed for other mass models. In fact, as mentioned above, the best accuracy is obtained for the WS+RBFoe approach, which is already very close to the chaos-related unpredictability limit for the calculation of nuclear masses~\cite{Barea2005PRL}.

\section{Summary an perspectives}

In this work, the RBF approach is extended to include the odd-even effects by separately training the RBF for the four groups of nuclei with different odd-even parities of ($Z$, $N$), i.e., the e-e, e-o, o-e, and o-o nuclei. Taking nine widely used nuclear mass models as examples, we found that the RBFoe approach can significantly improve the nuclear mass predictions. The rms deviations of all these models with respect to the known mass in AME2012 are reduced to less than $300$~keV. In particular, $135$~keV is obtained for the WS4+RBFoe approach, which is very close to the chaos-related unpredictability limit ($\sim 100$ keV). Moreover, the description of single-nucleon separation energies, $S_n$ and $S_p$, is better reproduced by the RBFoe approach. We also attempt to analyze the differences of mass predictions to the known data with their Fourier amplitudes, from which the regularity of mass differences for the original mass models and Model+RBF approaches is clearly found. However, no dominant frequency components are observed for the RBFoe approach, which again verifies the success of the RBFoe in reproducing the whole nuclear mass surface.

For further improving nuclear mass predictions, it is important to map the Fourier amplitude from the frequency region to the nuclear region. From Eq.~(\ref{Eq:Fk}), it is clear that the Fourier amplitude at a certain frequency corresponds to a sum on all nuclear mass differences weighted by exponential functions, so a certain frequency cannot easily map to a specific nuclear region. The wavelet transform provides a useful tool for this mapping, which might provide some indications for the further improvement of nuclear mass predictions. Therefore, it is interesting to analyze mass differences with the wavelet transform and investigations along these lines are in progress.

\section*{Acknowledgements}

This work was partly supported by the National Natural Science Foundation of China (Grants No. 11205004, No. 11105010, No. 11035007, and No. 11475014), the 211 Project of Anhui University under Grant No. 02303319-33190135, the Fundamental Research Funds for the Central Universities, and the RIKEN iTHES project.




\end{CJK*}
\end{document}